\documentclass[%
 reprint,
 amsmath,amssymb,
 aps,
]{revtex4-2}

\usepackage{graphicx}
\usepackage{dcolumn}
\usepackage{bm}
\usepackage[T1]{fontenc}
\usepackage{helvet}

\usepackage{physics}
\usepackage{subfigure}
\newtheorem{theorem}{Theorem}
\usepackage{bbold}
\usepackage{hyperref}
\usepackage{xcolor}
\usepackage{tikz}
\usetikzlibrary{positioning}

\usepackage{comment}
\newcommand{\sigvec }{\vec \sigma}

\begin{document}

\preprint{APS/123-QED}

\title{Designing dynamically corrected gates robust to multiple noise sources using geometric space curves}

\author{Hunter T. Nelson$^{**}$, Evangelos Piliouras$^{**}$, Kyle Connelly, and Edwin Barnes}
 \email{efbarnes@vt.edu}
 \thanks{\\\hspace*{-2ex}$^{**}$ These authors contributed equally.}
\affiliation{Department of Physics, Virginia Tech, Blacksburg, VA 24061, USA}
\affiliation{Virginia Tech Center for Quantum Information Science and Engineering, Blacksburg, VA 24061, USA}

\date{\today}

\begin{abstract}
Noise-induced gate errors remain one of the main obstacles to realizing a broad range of quantum information technologies. Dynamical error suppression using carefully designed control schemes is critical for overcoming this challenge. Such schemes must be able to correct against multiple noise sources simultaneously afflicting a qubit in order to reach error correction thresholds. Here, we present a general framework for designing control fields that simultaneous suppress both noise in the fields themselves as well as transverse dephasing noise. Using the recently developed Space Curve Quantum Control formalism, in which robust quantum evolution is mapped to closed geometric curves in a multidimensional Euclidean space, we derive necessary and sufficient conditions that guarantee the cancellation of both types of noise to leading order. We present several techniques for solving these conditions and provide explicit examples of error-resistant control fields. Our work also sheds light on the relation between holonomic evolution and the suppression of control field errors.
\end{abstract}

\maketitle

\section{Introduction}

Quantum information technologies require control schemes that satisfy both reachability and robustness criteria. Reachability refers to the set of achievable unitary evolutions under a given control scheme \cite{D'Alessandro2021}, while robustness refers to noise insensitivity, which is critical for reaching quantum error correction thresholds~\cite{Ng2011}. In recent years, the field of quantum optimal control has put forward several techniques for implementing a desired target evolution or logical gate operation while maintaining a certain level of noise robustness, including both analytical and numerical methods \cite{Biercuk2009,Khodjasteh2012,Low2014,Merrill2014,Brown2004,Sar2012,Khodjasteh2005,Green2013,Khodjasteh2009,Viola2003,Khodjasteh2010,Zhang2008,Viola1998,Constantin2010,Koch2016,Google2019,GIANNELLI2022128054,Koch2022,Damme2021,Ansel2021,KHANEJA2005296,Miller_2022,Chingiz2014,Khodjasteh2009Bounded}. While purely numerical approaches have been shown to be quite powerful in many contexts, with their convergence typically guaranteed, supplementing these with analytical information about the control landscape can further boost performance and potentially lead to globally optimal solutions.

Holonomic quantum computation \cite{guo2022optimizing,zanardi1999holonomic,wilczek1984appearance,berry2009transitionless,solinas2004robustness,sjoqvist2015geometric,sjoqvist2016conceptual,xu2012nonadiabatic,gungordu2014non,chen2020observable} is one example of an analytical technique that has been proposed to design robust control schemes. 
In this approach, gates are realized via the accumulation of geometric phases under holonomic evolution. Since such phases depend only on geometric properties of the path traversed in Hilbert space, and not on the rate of traversal, the resulting gates should be insensitive to noise that leaves the path invariant \cite{de2003berry,zhu2005geometric}. Evidence of such robustness against certain types of control field noise has been shown in both theoretical \cite{carollo2003geometric,de2008geometric,wang2007nonadiabatic,thomas2011robustness,liang2016proposal,chen2018nonadiabatic,liu2019plug,chen2020high,pachos2001quantum} and experimental work \cite{berger2013exploring,Zhou_PRL2017,kleissler2018universal,xu2020experimental}. However, whether or not these methods really provide robustness to noise across a broad range of practical settings has been a matter of contention, and several explicit examples of holonomic gates that do not exhibit the desired robustness are known~\cite{zhu2005geometric,nazir2002decoherence,blais2003effect,solinas2004robustness,carollo2003geometric,dajka2008bifurcations,ota2009composite,sjoqvist2012non,Zhuang_2022}. It was later found in Ref.~\cite{Colmenar2021} that for a large class of holonomic gates, it is possible to construct an equivalent non-holonomic gate with the same noise sensitivity, suggesting that even when robustness is achieved, holonomy may not be the underlying cause. On the other hand, it was shown in Ref.~\cite{colmenar2022efficient} that, under certain conditions, single-qubit gates that are simultaneously robust to multiplicative control field noise and transverse dephasing noise are necessarily holonomic. 

Another analytical approach to building noise-robust gates is through the Space Curve Quantum Control (SCQC) formalism ~\cite{Junkai2019,zeng2019geometric,buterakos2021geometrical,Zhuang_2022,zeng2018general,Bikun2021,barnes2022dynamically}. This method makes use of geometric space curves to design gates that are robust to transverse dephasing noise. Because this method provides access to the entire solution space of robust pulses, it offers the highest possible degree of flexibility in finding control waveforms that conform to experimental requirements. Most importantly, this approach does not rely on a conjecture---robustness is guaranteed by construction. However, so far SCQC has only been applied to transverse dephasing noise, while in most qubit platforms, including quantum dot spin qubits \cite{arif2021exploring,penthorn2019two,connors2022charge,culcer2012valley,mi2018landau}, superconducting transmons \cite{valery2022dynamical,houck2009life,metcalfe2007measuring}, and trapped ions \cite{talukdar2016implications,turchette2000decoherence}, control field noise is of comparable importance. Ref.~\cite{Dong2021-Dogs} took a first step in addressing this issue by showing how to incorporate holonomy into the SCQC formalism, yielding holonomic gates with guaranteed insensitivity to transverse noise. However, this approach is still susceptible to the usual uncertainties regarding whether holonomic evolution is robust to control field errors, leaving open the question of whether it is possible to guarantee the simultaneous cancellation of both transverse dephasing errors and control field errors in quantum gates.

In this work we present a general method for constructing single-qubit gates that are simultaneously robust to both transverse dephasing noise and multiplicative control field noise using the SCQC formalism. We do this in the context of a quasistatic noise model that is widely applicable across multiple qubit platforms, and show how to use this method to obtain waveforms that generate desired noise-robust logic gates. Suitable control waveforms are obtained by designing space curves that satisfy certain constraints and then computing their curvatures. We show that in addition to being closed, which guarantees the cancellation of transverse noise errors, the derivative of the curve must also satisfy a zero-area condition to ensure control field errors are also suppressed. We prove that these conditions are necessary and sufficient in that any robust pulse will be associated with curves satisfying these conditions. We also give several methods for constructing explicit examples of curves obeying these criteria, and we present a mathematical theorem that allows one to easily check whether a given curve derivative integrates to a closed curve. Additionally, we demonstrate how the freedom to switch between the dynamical and geometric phase without changing the robustness of gates, as first discussed in Ref.~\cite{Colmenar2021,colmenar2022efficient}, manifests in the SCQC formalism.

The paper is organized as follows. In Sec. II we utilize the SCQC formalism to derive space curve conditions that facilitate the construction of gates simultaneously robust to transverse noise and multiplicative control field noise. We also compare and contrast these conditions with the parallel transport condition associated with holonomic evolution. In Sec. III we discuss general methods for constructing space curves that yield noise-robust gates and compare the performance of gates generated using this approach. It is here that we also present a theorem that allows one to check if a given tangent curve produces a corresponding closed curve. We conclude in Sec.~\ref{sec:conclusion}.
\section{Noise cancellation conditions, space curves, and relation to holonomic evolution}
In this section, we consider a general single-qubit Hamiltonian simultaneously subject to two types of noise, one additive, the other multiplicative. Although the SCQC formalism has been extended to include the cancellation of time-dependent noise~\cite{Bikun2021}, in this work we focus solely on the case of quasistatic noise, which is pervasive in solid state qubits, where control time scales are fast compared to noise fluctuations~\cite{Dial_PRL13,OMalley_PRApplied15,Martins_PRL16,Hutchings_PRApplied17}. Since in most qubit platforms, the bulk of the noise is concentrated at low frequencies, the quasistatic model is already sufficient for designing high-fidelity control schemes. While further refinements can be achieved by accounting for noise coloration, we leave this to future work. Here, our goal is to derive conditions on space curves that guarantee the simultaneous cancellation of both types of quasistatic noise. In this section, we also examine the connection between these noise cancellation conditions and the conditions that define holonomic evolution.

\subsection{Noise cancellation conditions}
We begin with a three-field control Hamiltonian of the form: 
\begin{align}
\begin{split}
      H_0(t)= \frac{\Omega(t)}{2}(\cos\Phi(t)\sigma_x + \sin\Phi(t) \sigma_y)+ \frac{\Delta(t)}{2} \sigma_z,
\end{split}
\end{align}
where we refer to $\Omega(t),\Phi(t),\Delta(t)$ as the driving, phase, and detuning fields, respectively, and $\sigma_x$, $\sigma_y$, $\sigma_z$ are Pauli matrices.
We consider multiplicative errors in $\Omega(t)$ and additive errors in $\Delta(t)$:
\begin{align}
    \Omega(t) &\rightarrow (1+\epsilon)\Omega(t), \\
    \Delta(t) &\rightarrow \Delta(t) + \delta_z,
\end{align}
where $\epsilon$ and $\delta_z$ are unknown, stochastic noise parameters that are assumed to be small and constant during the evolution.
This model captures the common situation in which noise causes a slow, random rescaling of the driving field, as occurs for instance in exchange pulses in quantum dot spin qubits subject to charge noise~\cite{Dial_PRL13}. Additive fluctuations in $\Delta(t)$ are a widely used model of dephasing noise in qubit energy levels, where the dephasing time $T_2^*$ is set by the width of the distribution from which $\delta_z$ is sampled~\cite{Martins_PRL16}. 

In order to quantify the deviation away from the ideal evolution caused by $\epsilon$ and $\delta_z$, it helps to switch to the interaction picture defined by $U_0(t)$, the evolution operator generated by $H_0(t)$. The Magnus expansion \cite{BLANES2009151} of the interaction picture evolution operator is then controlled by the small parameters $\epsilon$ and $\delta_z$. At first order we have
\begin{align}
    U_I(t) \approx e^{-i\Pi_1(t)},
\end{align}
with
\begin{align}
\begin{split}
     \Pi_1(t) = \int_0^t dt' H_I(t') = 
     \frac{\delta_z}{2}\int_0^t dt' U_0^\dag(t')\sigma_z U_0(t')\\+
    \frac{\epsilon}{2}\int_0^t dt'  U_0^\dag(t')\Omega(t')\,(\cos\Phi(t') \sigma_x +\sin\Phi(t')  \sigma_y)U_0(t') .
    \label{first-order-magnus-with-fields}
\end{split}
\end{align} 
Following \cite{Junkai2019}, we interpret the term proportional to $\delta_z$ as a curve in three-dimensional (3D) Euclidean space that we refer to as the ``space curve" or ``error curve" $\vec{r}(t)$:
\begin{equation}
    \vec{r}(t) \cdot \vec{\sigma} \equiv \int_0^t \dd{t'} U_0^\dag(t') \sigma_z U_0(t').
    \label{eq:space_curve}
\end{equation}
By construction, it then follows that 
cancelling transverse dephasing noise to first order in $\delta_z$ corresponds to ensuring that $\vec{r}(t)$ is a closed curve. To study these 3D space curves, we define an orthonormal frame called the Frenet-Serret frame \cite{diff-geom-book}, consisting of the tangent vector $\vec{T} \equiv \dot{\vec{r}}$, the normal vector $\vec{N} \equiv \dot{\vec{T}}/\norm{\dot{\vec{T}}}$, and the binormal vector $\vec{B} \equiv \vec{T} \times \vec{N}$. These vectors then satisfy the Frenet-Serret equations: 
\begin{align}
    \dot{\vec{T}} &= \kappa \vec{N}, \nonumber\\
    \dot{\vec{N}} &= -\kappa \vec{T} + \tau \vec{B}, \\
    \dot{\vec{B}} &= -\tau \vec{N}. \nonumber
\end{align}
The functions $\kappa$ and $\tau$ are the curvature and torsion of the curve, and via the Frenet-Serret equations they uniquely determine the curve up to rigid rotations, in an interval where $\kappa \neq 0$ (see Appendix \ref{Curve to fields mapping}). Once we find a closed space curve, we can find the corresponding control fields $\Omega$, $\Phi$, and $\Delta$ from the curvature $\kappa$ and torsion $\tau$ of the space curve:
\begin{align}
    \kappa &= \Omega, \\
    \tau &= \dot{\Phi} - \Delta.
    \label{torsion-field-eq}
\end{align}
We see that any closed space curve yields control fields that generate a quantum evolution that is insensitive to quasistatic transverse dephasing errors. Note also that $\Phi$ and $\Delta$ are not uniquely determined by the geometry of the space curve. 

The second term in Eq.~\eqref{first-order-magnus-with-fields} can also be written in terms of the space curve. From the definition of the space curve in Eq.~\ref{eq:space_curve} we see that:
\begin{align}
\begin{split}
     \dot{\vec T}(t) \cdot \sigvec  &= \frac{d}{dt}(U_0^{\dagger} \sigma_z U_0)\\
    &= i U_0^{\dagger}[H_0,\sigma_z] U_0\\
&=-2i \left(U_0^{\dagger} \sigma_z U_0 \right) U_0^{\dagger} \begin{pmatrix}
0 & \frac{\Omega(t)}{2} e^{-i \Phi(t)} \\
\frac{\Omega(t)}{2} e^{i \Phi(t)} & 0
\end{pmatrix} U_0\\
&=-2i ({\vec T}(t)\cdot \sigvec) U_0^{\dagger} \begin{pmatrix}
0 & \frac{\Omega(t)}{2} e^{-i \Phi(t)} \\
\frac{\Omega(t)}{2} e^{i \Phi(t)} & 0
\end{pmatrix} U_0,\\
\end{split}
\end{align}
which implies that
\begin{align}
\begin{split}
    \Omega(t)\,U_0^\dag(t)(\cos\Phi(t) \sigma_x &+\sin\Phi(t)  \sigma_y)U_0(t)= \\& = i({\vec T}(t) \cdot \sigvec) (\dot{\vec T}(t) \cdot \sigvec) \\ 
     &= -(\vec{T}(t) \times \dot{ \vec T}(t))\cdot \sigvec.
\end{split}
\label{omega-t-tdot}
\end{align}
Thus, we see that the leading-order errors from both types of noise can be expressed in terms of the tangent curve $\vec T(t)$, with Eq.~\eqref{first-order-magnus-with-fields} becoming
\begin{align}
      \Pi_1(t) = \frac{\sigvec}{2}\cdot \int_0^t dt' ( -\epsilon \,\,\vec{T} \times \dot{ \vec T} + \delta_z \vec{T}).
      \label{Magnus exp geomtric}
 \end{align}
A doubly robust qubit evolution ($U_I(T) \approx \mathbb{1})$ then requires that the following two conditions be simultaneously satisfied:
\begin{align}
    \int_0^T dt \,\vec{T} &= 0,
    \label{1st order magnus constraints1} \\
         \int_0^T dt \,(\vec{T} \times \dot{ \vec T})  &= 0.
      \label{1st order magnus constraints2}
\end{align}
The second condition is proportional to the area swept out by the projection of the tangent vector onto each plane. (Interestingly, it was found in Ref.~\cite{Junkai2019} that if the same condition is satisfied by the error curve itself, the dephasing error is suppressed to second order.) Therefore to cancel both types of error to first order, we must find a closed space curve $\vec{r}$ whose tangent vector $\vec{T}$ sweeps out zero area when projected onto any plane.

\subsection{Relation to the holonomic evolution}
\label{Geometric gate comparison}

Much of the previous work on designing robust quantum gates focused on making the evolution holonomic, meaning that the dynamical phase vanishes at the final gate time \cite{de2003berry,zhu2005geometric,Dong2021-Dogs}.
Here, we express the dynamical phase in terms of the tangent vector $\vec T$, and by comparing this to the noise-cancellation conditions, Eqs.~\eqref{1st order magnus constraints1} and \eqref{1st order magnus constraints2}, we show that holonomic evolution is neither necessary nor sufficient to guarantee robust evolution.

The dynamical phase is given by \cite{Aharonov_PRL87}
\begin{align}
\begin{split}
    \alpha_d(t) &= \int_{0}^{t} \bra{\psi(t')} H_0(t') \ket{\psi(t')} dt'\\ 
    &= \int_{0}^{t} \bra{\psi(0)} U_0^{\dagger}(t') H_0(t') U_0(t') \ket{\psi(0)}dt'.\\
    \end{split}
\end{align}
Defining the initial Pauli vector as $\vec P(0) = \bra{\psi(0)}\vec \sigma \ket{\psi(0)}$ and using Eq.~\eqref{omega-t-tdot}, the dynamical phase may then be written as:
\begin{align}\label{eq:dynamical_phase}
\begin{split}
     \alpha_d(t) &= \frac{1}{2} \int_{0}^{t} \bra{\psi(0)} (- \vec{T} \times \dot{ \vec T}+\Delta(t') \vec{T})\cdot \sigvec \ket{\psi(0)} \\
     &= \frac{\vec P(0)}{2} \cdot \int_{0}^{t} (-\vec{T} \times \dot{ \vec T}+\Delta(t') \vec{T})\, dt'.
     \end{split}
\end{align}
Despite the striking resemblance to Eq.~\eqref{Magnus exp geomtric}, cancelling the dynamical phase does not necessarily lead to error robustness.
To elaborate on this, we first observe from Eq.~\eqref{eq:dynamical_phase} that the dynamical phase involves a projection of the integral onto the Pauli vector of the initial state. Making this vanish requires either a special choice of the detuning field $\Delta(t)$ or a carefully designed tangent vector $\vec T$ in order to ensure the integral in Eq.~\eqref{eq:dynamical_phase} is orthogonal to $\vec P(0)$ at the final time.  Another approach one can take is to impose the parallel transport condition, which is tantamount to requiring that the integrand in Eq.~\eqref{eq:dynamical_phase} is orthogonal to $\vec P(0)$ at all times \cite{Dong2021-Dogs}. Regardless of which approach is taken to make $\alpha_d(T)$ vanish, it does not necessarily follow that Eqs.~\eqref{1st order magnus constraints1} and \eqref{1st order magnus constraints2} are satisfied. Therefore, the condition that the evolution be holonomic does not guarantee that either type of noise is canceled.

Conversely, it also holds that a robust evolution does not necessarily have to be holonomic. To see this, consider taking a holonomic evolution $U(t)$ and forming a new evolution $\Tilde{U}(t) = Z_{\Lambda(t)}U(t)Z_{-\Lambda(0)}$, where $Z_{\Lambda(t)}=e^{-i\frac{\Lambda(t)}{2}\sigma_z}$, and $\Lambda(t)$ is an arbitrary differentiable function. This new evolution has the initial condition $\tilde{U}(0) = \mathbb{1}$ and evolves under the effective Hamiltonian:
\begin{align}
    \tilde{H} &\equiv i \dot{\tilde{U}}\tilde{U}^\dagger = Z_\Lambda H
    Z_{\Lambda}^\dag + i \qty(\dv{t}Z_\Lambda) Z_\Lambda^\dagger \nonumber\\ 
    &= \frac{\Omega}{2}\qty(\cos\tilde{\Phi}\sigma_x + \sin\tilde{\Phi}\sigma_y) + \frac{\tilde{\Delta}}{2}\sigma_z,
\end{align}
where $\tilde{\Phi} \equiv \Phi + \Lambda$, $\tilde{\Delta} \equiv \Delta + \dot{\Lambda}$. Therefore changing the driving fields from $\qty{\Phi, \Delta} \rightarrow \qty{\tilde{\Phi}, \tilde{\Delta}}$ will implement the gate $\tilde{U}$ rather than $U$. Changing the control fields in this way leaves the curvature and torsion $\tau = \dot{\Phi} - \Delta$ invariant, and thus $\vec{\tilde{r}}$ can only differ from $\vec{r}$ by a rigid rotation. However, since the error Hamiltonian can be written as $H_I = \frac{1}{2}\qty(-\epsilon \vec{T}\times\dot{\vec{T}} + \delta_z\vec{T})\cdot\vec{\sigma}$, changing the phase and detuning fields in this way can only change $\tilde{H}_I$ and $\tilde{U}_I$ by a constant rotation, and so such a change does not affect the robustness of the evolution to any order. This does, however, change the dynamical phase, since $\Lambda(t)$, and hence $\tilde\Delta(t)$, can be chosen arbitrarily. A similar observation was made in Ref.~\cite{Colmenar2021} using the filter function formalism. Additionally we note here that $\tilde{U}(T)$ does not have to be equal to $U(T)$. By properly choosing $\Lambda(T)$ and $\Lambda(0)$, we can add arbitrary $z$-rotations before and after a robust gate without affecting its robustness.

One exception to the above analysis occurs in the case of a constant detuning field $\Delta$, where robustness to both types of noise does imply that the evolution is holonomic, as first pointed out in Ref.~\cite{colmenar2022efficient}. This is readily seen from the SCQC formalism by observing that if a space curve satisfies Eqs.~\eqref{1st order magnus constraints1} and \eqref{1st order magnus constraints2}, then from Eq.~\eqref{eq:dynamical_phase} it immediately follows that $\alpha_d(T)=0$. Note however that the dynamical phase is only guaranteed to vanish if $\Delta$ is constant. Transformations $Z_\Lambda$ that change the detuning field from one constant value to another, i.e., ones for which $\Lambda(t)$ is linear in $t$, will preserve the holonomic condition.

\section{Construction of Doubly robust evolutions}
We now proceed to construct several classes of curves satisfying the conditions given in Eqs.~\eqref{1st order magnus constraints1} and \eqref{1st order magnus constraints2}. These conditions ensure that the driving pulses produced from these curves are doubly robust to first-order multiplicative driving field noise and additive transverse dephasing noise. We present three different methods for constructing curves that satisfy these conditions. The first method utilizes an ansatz consisting of even and odd parity space curve components comprised of trigonometric functions with frequencies fixed such that both robustness conditions are satisfied. We refer to such curves as "parity curves". The second method accomplishes the cancellation of errors by instead utilizing an ansatz for the tangent curve for which Eq.~\eqref{1st order magnus constraints2} is enforced by symmetry.
Choosing parameters equal to Bessel function roots guarantees closure of the space curve so that Eq.~\eqref{1st order magnus constraints1} is also satisfied, yielding "Bessel curves". The third method consists of constructing the tangent curve on a sphere such that it traces out "tilted circles" that sweep zero area while also containing the origin in its convex hull. We prove that the latter condition ensures the corresponding space curve is closed, and so both Eqs.~\eqref{1st order magnus constraints1} and \eqref{1st order magnus constraints2} are again satisfied.

For each example curve presented below, we confirm the robustness of the resulting gate by numerically simulating the evolution in the presence of both types of errors and computing the gate infidelity $\mathcal{I}$ using the definition introduced in Ref.~\cite{PEDERSEN200747} (see also Ref.~\cite{Dong2021-Dogs}).

\subsection{Parity Curves}
First, we consider the approach that utilizes the parity and periodicity of trigonometric functions. This class of curves can be written in the form 
\begin{equation}
    \vec{r}(\lambda)=f_x(\omega_x\lambda)\hat x+f_y(\omega_y\lambda)\hat y+f_z(\omega_z\lambda)\hat z,
\end{equation}
where each function $f_i(\omega_i \lambda)$ is periodic with period $2\pi/\omega_i$ and either even or odd in $\lambda$. Parameterizations of this sort make it straightforward to impose symmetries in the space curve, and this in turn can make it easier to satisfy the noise-cancellation conditions. In particular, curves whose components are all odd or all even functions satisfy these conditions. These "parity curves" are related to the trigonometric curves considered in the literature on the differential geometry of curves \cite{animov2001differential}. The periodicity of such functions guarantees the curve is closed, provided we choose all the ratios of the frequencies to be rational numbers so that a least common multiple always exists. The parity property of the trigonometric functions ensures the curves have vanishing projected areas for both the space and tangent curves. For instance, if we start with a curve whose components are of the even type, then $\dot{\vec{r}}$ will be comprised of odd functions, and $\ddot{\vec{r}}$ will be comprised of even functions. Therefore the integral of the cross product, Eq.~\eqref{1st order magnus constraints2}, will only contain odd functions and will therefore vanish over the period of the curve. 

\begin{figure}
    \centering
    \includegraphics[width=1.1\columnwidth,
    trim = {10cm 1cm 10cm 0},clip
    ]{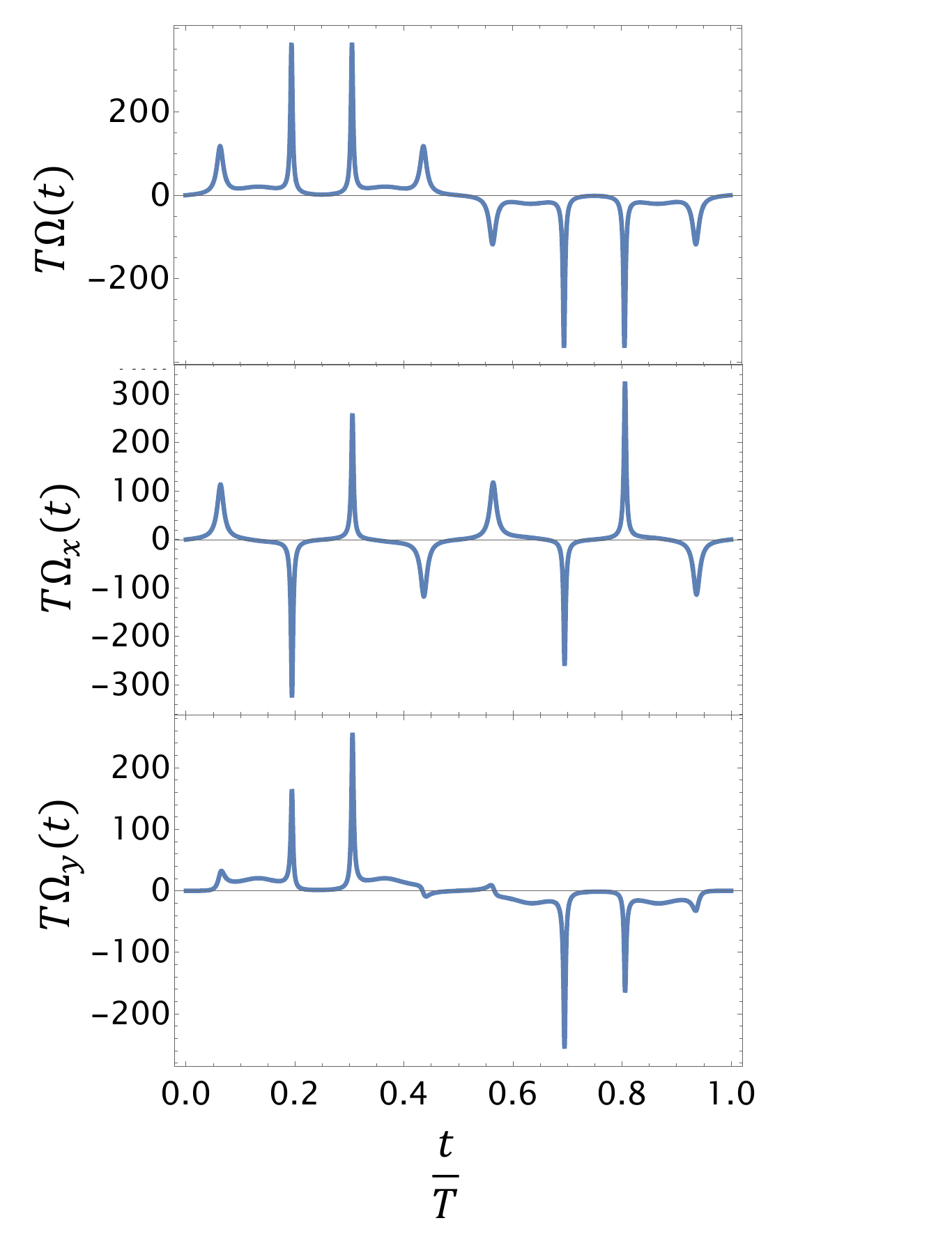}
    \caption{Control fields that generate an identity gate robust to both control field noise and transverse dephasing noise obtained using the parity curve given in Eq.~\eqref{robust curve}. Here, $\Omega_x=\Omega\cos\Phi$, $\Omega_y=\Omega\sin\Phi$, $\Delta=0$, and $T$ is the gate time.}\label{Fourier control fields}
\end{figure}

The following is an example of a parity curve, where we choose each component to be odd: 
\begin{align}
    \vec{r}(\lambda) &= \sin(\lambda/2) \hat{x} + \sin(\lambda) \cos^2(\lambda) \hat{y}+\sin(\lambda) \hat{z}. \label{robust curve}
\end{align}
Here, $\lambda\in[0,4\pi]$. Before we can extract a pulse from this curve, we must first switch to the arclength parameterization $t$ defined by $\norm{\frac{d}{dt}\vec{r}(\lambda(t))}^2=1$ and such that $t\in[0,T]$, where $\lambda(T)=4\pi$. It follows immediately that $\vec{r}(T)=\vec{r}(\lambda(T))=0$, and because of the built-in parity symmetry, the corresponding tangent curve $\vec{T}(t)=\dot{\vec{r}}(t)$ satisfies Eq.~\eqref{1st order magnus constraints2}, implying that both types of noise are cancelled. In fact, this curve also has vanishing projected areas, $\int_0^{T} dt \,\,\vec{r}(t)\times\vec{T}(t)=0$, so transverse dephasing noise is actually cancelled up to second order in this example. The control fields obtained from the curvature and torsion of $\vec{r}(t)$ using Eq.~\eqref{torsion-field-eq} are shown in Fig.~\ref{Fourier control fields}.

\begin{figure*}
\includegraphics[width=\textwidth,
trim={0 2cm 0 0},clip]{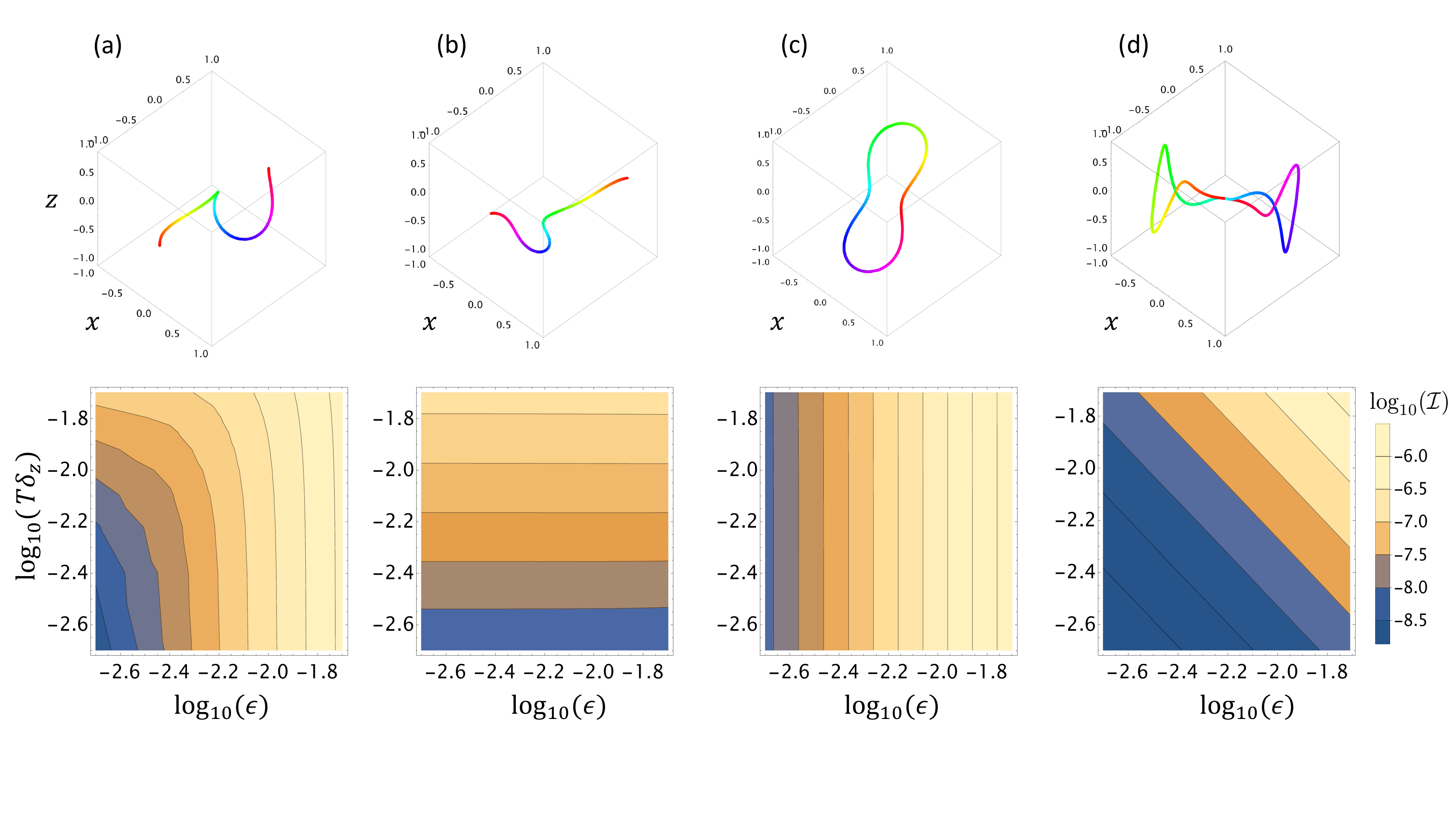}
\caption{Four space curves (top row) that correspond to identity gates and the corresponding gate infidelities $\mathcal{I}$ (bottom row) as a function of control field noise ($\epsilon$) and transverse dephasing noise ($\delta_z$).
    (a) The space curve defined in Eq.~\eqref{non-robust} which does not satisfy either of the noise-cancellation constraints, Eqs.~\eqref{1st order magnus constraints1} and \eqref{1st order magnus constraints2}.
     (b) The space curve defined in Eq.~\eqref{non-closed} which satisfies Eq.~\eqref{1st order magnus constraints2} but not Eq.~\eqref{1st order magnus constraints1}.
    (c) The space curve defined in Eq.~\eqref{non-zero area} which satisfies Eq.~\eqref{1st order magnus constraints1} but not Eq.~\eqref{1st order magnus constraints2}.
    (d) The doubly robust "parity" space curve defined in Eq.~\eqref{robust curve} which satisfies both Eqs.~\eqref{1st order magnus constraints1} and \eqref{1st order magnus constraints2}.
    }\label{Fourier comparison}
\end{figure*}

To demonstrate the improved performance of the gate generated by this curve, we compare its infidelity against that of three other curves for which at least one of the noise-cancellation conditions, Eqs.~\eqref{1st order magnus constraints1} and \eqref{1st order magnus constraints2}, is not satisfied. These three curves have the following tangent curves:
\begin{align}
    \dot{\vec{r}}(t) &= \sin(t)\cos(t) \hat{x} + \sin^2(t) \hat{y} + \cos(t) \hat{z}, \label{non-robust}\\
    \dot{\vec{r}}(t) &= \frac{1}{4}(\sqrt{2} \cos(2t)-2\cos(t)) \hat{x}\nonumber\\
    &- \frac{1}{4}(\sqrt{2} \sin(2t)+2\sin(t)) \hat{y}\nonumber \\
    &+ \frac{1}{2}\sqrt{\sqrt{2} \cos(3t)+\frac{5}{2}} \hat{z}, \label{non-closed}\\
    \dot{\vec{r}}(t) &=\sin(t)\cos(2t) \hat{x} +\sin(t)\sin(2t)  \hat{y}+ \cos(t) \hat{z},  \label{non-zero area}
\end{align}
where $t\in[0,2\pi]$ and we have set $T=1$ to simplify the notation.
Here, Eq.~\eqref{non-robust} is a non-robust curve that breaks both conditions \eqref{1st order magnus constraints1} and \eqref{1st order magnus constraints2}, Eq.~\eqref{non-closed} has vanishing area but generates a non-closed curve, and Eq.~\eqref{non-zero area} does not sweep zero area but does generate a closed curve. The space curves obtained by integrating these tangent curves are shown in Fig.~\ref{Fourier comparison}(a)-(c). We also note that Eq.~\eqref{non-closed}, which is built using the technique given in Ref.~\cite{weiner1977closed}, is a less trivial example of a tangent curve with vanishing area that does not use the parity property built into Eq.~\eqref{robust curve}. The fully robust curve of Eq.~\eqref{robust curve} is shown in Fig.~\ref{Fourier comparison}(d). All four of these curves produce identity gates. 

The gate infidelities corresponding to all four of the above curves are shown in Fig.~\ref{Fourier comparison} as a function of the strengths of both types of noise. Figure~\ref{Fourier comparison}(a) clearly shows the reduction in performance that results when the space curve breaks both conditions, Eqs.~\eqref{1st order magnus constraints1} and \eqref{1st order magnus constraints2}, with the infidelity growing rapidly as both noise strengths are increased. On the other hand, Figs.~\ref{Fourier comparison}(b) and \ref{Fourier comparison}(c) both exhibit a clear robustness against one type of noise depending on which noise-cancellation condition is satisfied. Finally, Fig.~\ref{Fourier comparison}(d) shows a marked improvement in suppressing both types of noise, where now both noise-cancellation conditions are satisfied. The infidelity in this case in fact scales better than the expected $\epsilon^2$, $(T \delta_z)^2$ scaling. The improved scaling in $T \delta_z$ can be understood from the fact that the projected areas of the space curve all vanish, as noted above. The $\epsilon$ scaling suggests that the $\epsilon^2$ term in the Magnus expansion also vanishes for this example; this in turn may be a consequence of the parity symmetry. Further investigation of the higher-order terms of the Magnus expansion would be needed to confirm this. Finally, we note that although the use of parity curves makes it easy to satisfy both noise-cancellation constraints at the same time, the control fields in this example as shown in Fig.~\ref{Fourier control fields} may be hard to implement in practice due to their sharp (although non-singular) features. In the subsequent sections, we present two other methods of constructing curves that satisfy both noise-cancellation constraints while also yielding more experimentally friendly pulse shapes.


\subsection{Bessel curves}

\begin{figure*}
\includegraphics[width=\textwidth]{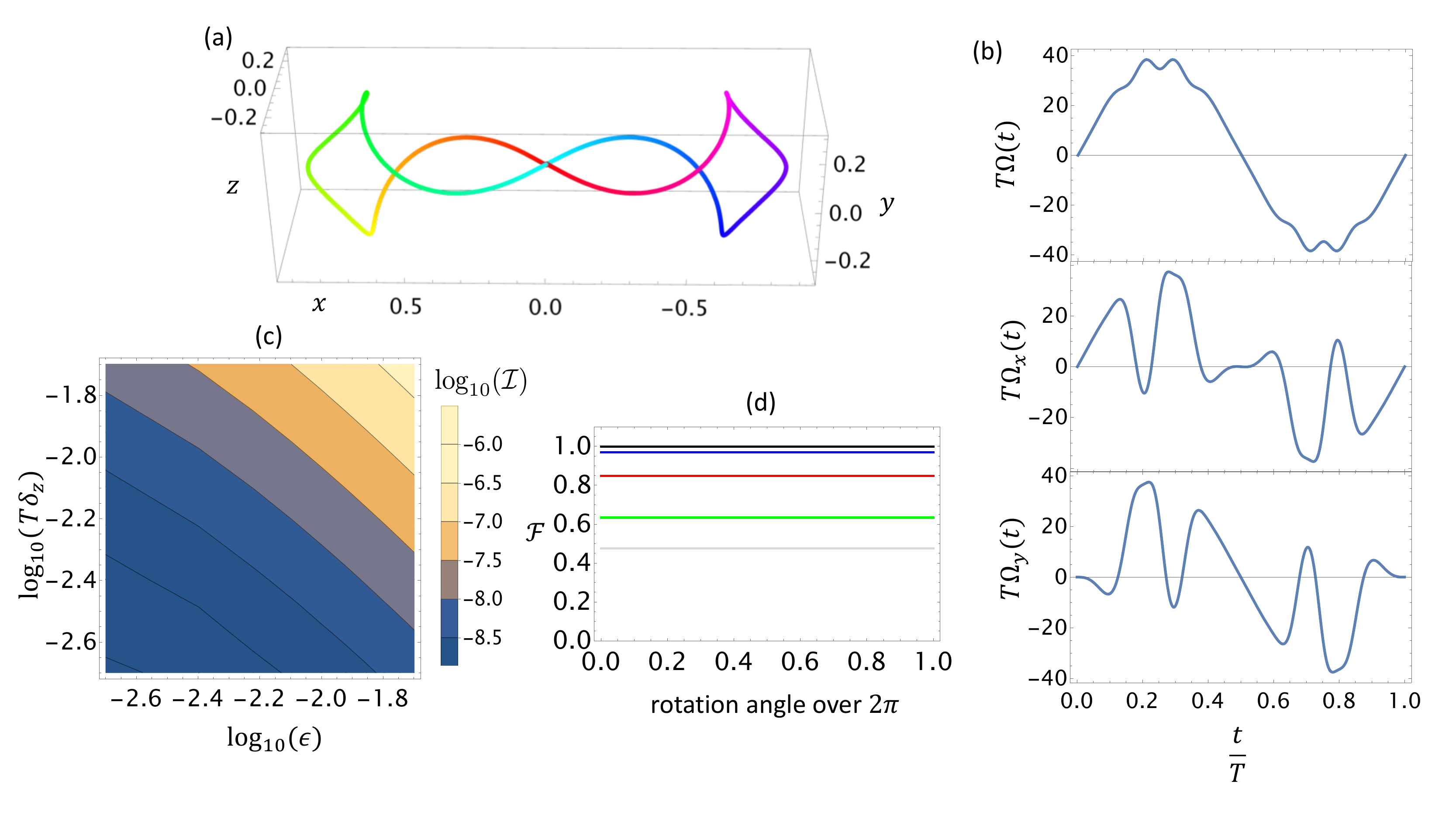}
\caption{(a) The "Bessel curve" obtained from Eqs.~\eqref{sphere curve} and \eqref{Bessel_theta} with $x_i=5.5201$, $q=0.5660$. (b) Control fields that implement a doubly robust identity gate obtained from the curvature and torsion of the Bessel curve in (a). Here, $\Omega_x=\Omega\cos\Phi$, $\Omega_y=\Omega\sin\Phi$, $\Delta=0$, and $T$ is the gate time. (c) The infidelity of $z$-rotations as a function of the strengths of transverse dephasing noise $(\delta_z)$ and multiplicative driving field noise $(\epsilon)$. The infidelity for each value of $\delta_z$ and $\epsilon$ is averaged over the uniform distribution $[0,2\pi]$ of $z$-rotation angles.  (d) The fidelity $\mathcal{F}$ versus $z$-rotation angle (divided by $2\pi$). Each line corresponds to different values of $\epsilon$ and $\delta_z$ (both have the same value along each line). Starting from the top, the error values are $\delta_z,\epsilon=0$, $0.1$, $0.2$, $0.3$, and $0.4$, respectively.}
 \label{Bessel curve}
\end{figure*}
Our second technique for constructing curves that satisfy Eqs.~\eqref{1st order magnus constraints1} and \eqref{1st order magnus constraints2} starts by defining the following ansatz for the normalized tangent curve:
\begin{align}
    \vec T(\theta(t)) = (\cos(q\theta)\sin\theta,\sin(q\theta)\sin\theta,\cos\theta), \label{sphere curve}
\end{align}
where $q$ is a proportionality constant between the azimuthal and polar angles. This formulation provides a curve that is already expressed in its own arclength parameterization and which is solely controlled by a function $\theta(t)$. This ansatz makes the pulse error constraint particularly simple:
\begin{align}
    \int_0^T dt \,(\vec{T} \times \dot{ \vec T}) = \int_{\theta(0)}^{\theta(T)} d\theta \,(\vec{T} \times  \frac{\partial\vec {T}}{\partial \theta})=0,
\end{align}
where examination of the third component forces us to require
\begin{equation}
    \theta(0)=\theta(T).\label{vanish curve tantrix cond}
\end{equation}
This one boundary constraint is equivalent to the vanishing area condition, Eq.~\eqref{1st order magnus constraints2}, along all 3 projections. For the curve to be closed, we additionally need 3 real integrals to vanish. These integrals come from plugging Eq.~\eqref{sphere curve} into Eq.~\eqref{1st order magnus constraints1}. However, to simplify the process of finding suitable functions $\theta(t)$, we will upgrade these to 3 complex integral constraints:
\begin{align}
   \int_0^T e^{i\theta(t)} dt  &= 0 \label{closed curve tantrix cond1} , \\
    \int_0^T e^{i\theta(t)(1\pm q)} dt &= 0,\label{closed curve tantrix cond2} 
\end{align}
where Eq.~\eqref{closed curve tantrix cond2} should be understood as two separate constraints, one for each choice of the sign in front of $q$.
Although Eqs.~\eqref{closed curve tantrix cond1} and \eqref{closed curve tantrix cond2} are generally stronger constraints than Eq.~\eqref{1st order magnus constraints1}, these complex constraints have the advantage that they can be solved approximately using Bessel functions, as we explain in more detail below. 

We note in passing that if we choose $q=0$, then Eq.~\eqref{closed curve tantrix cond2} becomes redundant, and the one independent integral constraint that remains, Eq.~\eqref{closed curve tantrix cond1}, coincides with the closed-curve constraint for plane curves \cite{zeng2018general}. This is to be expected, since setting $q=0$ in Eq.~\eqref{sphere curve} restricts the tangent curve, and hence the space curve, to the $xz$ plane. In this case, $\dot\theta(t)$ can be interpreted as the curvature of the plane curve \cite{barnes2022dynamically}, and Eq.~\eqref{1st order magnus constraints1} is equivalent to  Eq.~\eqref{closed curve tantrix cond1}. The zero-area constraint on the tangent curve, Eq.~\eqref{1st order magnus constraints2}, remains equivalent to Eq.~\eqref{vanish curve tantrix cond}.

For any other value of $q\ne0$,
Eq.~\eqref{closed curve tantrix cond2} imposes an independent constraint on the space curve. The magnitude of the curvature of the space curve is given by
\begin{align}
    |\kappa(t)| = \sqrt{q^{2} \sin^{2}{\left(\theta{\left(t \right)} \right)} + 1}\, |\dot{\theta}(t)|.
\end{align}
We see that if we impose $\dot\theta(0)=0=\dot\theta(T)$, the resulting pulse envelope $\Omega(t)=\kappa(t)$ will start and end at zero as should be the case for a smooth pulse. This condition and Eq.~\eqref{vanish curve tantrix cond} are both satisfied by the following ansatz: 
\begin{align}\label{Bessel_theta}
    \theta(t) = x_i \cos(\frac{2 \pi}{T}t),
\end{align}
where $x_i$ is a real constant. We refer to the space curve obtained from this choice of $\theta(t)$ as a "Bessel curve". Inserting this ansatz into Eqs.~\eqref{closed curve tantrix cond1} and \eqref{closed curve tantrix cond2} and comparing the results to the integral representation of the Bessel function of the first kind  \cite{Jackson1999}, we see that all the space curve constraints are satisfied by choosing $x_i$ and $(1\pm q)x_i$ to be Bessel function zeros: 
\begin{align}
    J_0(x_i) = 0, \\
    J_0((1\pm q)x_i) = 0.\label{J0_q}
\end{align}
In the case of a plane curve ($q = 0$), the resulting evolution generated by the sinusoidal $\Omega(t)$ is robust to pulse errors since Eq.~\eqref{vanish curve tantrix cond} holds; more surprisingly, we see that there exist particular pulse amplitudes $x_i$ for which dephasing noise is also suppressed. For more general values of $q\ne0$, we have a 3D space curve, and we must choose $q$ so that both $(1+q)x_i$ and $(1-q)x_i$ are also Bessel function zeros. Although we cannot in general find values of $q$ for which these quantities are both exact zeros, we can make them approximate zeros by finding a $q$ that minimizes $[(1-q)x_i - x_{i-1}]^2 + [(1+q)x_i - x_{i+1}]^2$. For three consecutive exact Bessel zeros $x_{i-1}$, $x_i$, $x_{i+1}$, this function is approximately minimized when
\begin{align}
    q = \frac{x_{i+1}-x_{i-1}}{2 x_i}. \label{central diff}
\end{align}
The parameters $x_i$ and $q$ together give us discrete control over the smoothness of the space curve and, hence, the bandwidth of the resulting control field. The curve and associated control fields for the choice $x_i=5.5201$, $q=0.5660$ are shown in Figs.~\ref{Bessel curve}(a) and \ref{Bessel curve}(b), respectively. In Fig.~\ref{Bessel curve}(c), we confirm the expected insensitivity of the resulting $z$ gates to both transverse dephasing noise and multiplicative control field noise. Different $z$-rotations are constructed by adjusting the gauge choice for $\Phi$ and $\Delta$ as discussed in the previous section. Figure~\ref{Bessel curve}(d) shows that this robustness persists across all possible $z$-rotations.
It is noteworthy that in this example, the particular space curve also happens to have vanishing projected areas, and so therefore second-order dephasing errors are also suppressed.

\subsection{Tilted circles}

\begin{figure*}
\includegraphics[width=\textwidth]{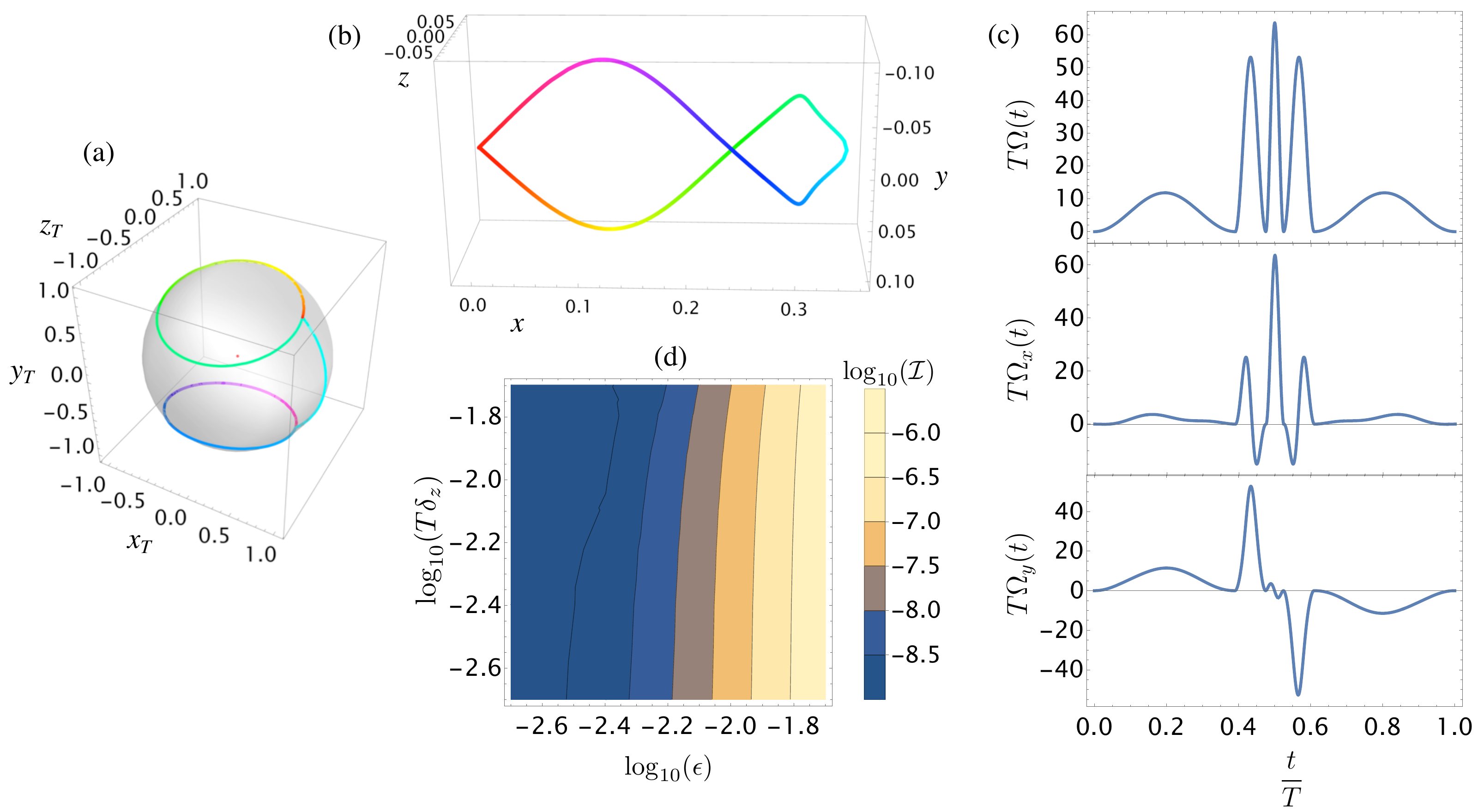}
\caption{(a) The "tilted circles" tangent curve, shown here for $\theta = \pi/2$. The hue of the curve indicates the manner in which it is traced. We also show the origin (in red) to show that it is contained in the convex hull of this curve. (b) A closed space curve generated from the tilted circle tangent curve in (a). The hue of this curve matches that of the tangent curve, and shows the speed at which different sections of the tangent curve are traversed. (c) Control fields for a doubly robust $R(\frac{\pi}{2}X)$ gate extracted from the curvature and torsion of the curve in (b). To achieve the desired $z$-rotation angle, a constant detuning field is chosen appropriately. (d) Infidelity of the tilted curve-based $R(\frac{\pi}{2}X)$ gate   versus transverse dephasing noise strength $(\delta_z)$ and multiplicative driving field noise strength $(\epsilon)$.}\label{tilted curve}
\end{figure*}

Another very general strategy for constructing curves that satisfy the robustness conditions comes from the observation that the integral giving the area swept out by the tangent curve is independent of the parameterization of that curve:
\begin{equation}
    \int_0^{T} \vec{T} \times \dot{\vec{T}} \dd{t} = \int \vec{T} \times \dd{\vec{T}}.
\end{equation}
Thus we can start by drawing a tangent curve that sweeps out zero area, and then try to find a parameterization such that the space curve is closed. Let $s$ be the arclength of $\vec{T}$, i.e., $\norm{\dv*{\vec{T}}{s}} = 1$. Then after designing a tangent curve $\vec{T}(s)$ that sweeps out zero area, we must find a parameterization $s(t)$ that gives a closed space curve upon integrating $\vec{T}(s(t)) \equiv \vec{T}(t)$.\\
\\
The curvature $\kappa$ and torsion $\tau$ of $\vec{r}$ are related to $\vec{T}(t)$ as follows:
\begin{equation}
  \kappa = \norm{\dot{\vec{T}}} = \dv{s}{t},
\end{equation}
\begin{align}
  \tau &= -\vec{N} \cdot \dot{\vec{B}} = -\vec{N} \cdot \dv{t}\qty(\vec{T}\times\vec{N}) =  -\frac{1}{\kappa} \dot{\vec{T}} \cdot \dv{t} \qty(\vec{T} \times \frac{1}{\kappa} \dot{\vec{T}}) \nonumber\\
  &= -\frac{1}{\kappa}\dv{t}\vec{T} \cdot \vec{T} \times \kappa\qty(\frac{1}{\kappa}\dv{t})^2\vec{T} = \kappa \vec{T} \cdot \dv{\vec{T}}{s} \times \dv[2]{\vec{T}}{s}.
\end{align}
The vector triple product above is the geodesic curvature of the tangent curve $\kappa_{g,T}$, so this relationship can be written as $\tau/\kappa = \kappa_{g,T}$.

Not every tangent curve can be reparameterized to give a closed space curve however. For instance, in Eq. \eqref{non-closed} the $z$ component of $\vec{T}$ is always positive, and so the $z$ component of $\vec{r}$ will be non-decreasing regardless of how $\vec{T}$ is parameterized. Fortunately we find a simple visual criterion that can be used to determine if a given $\vec{T}$ curve can yield a closed space curve: 

\begin{theorem}
    A tangent curve $\vec{T}(s)$ can generate a closed space curve if and only if the convex hull of $\vec{T}(s)$ contains the origin.
    \label{thm:conv_hull}
\end{theorem}

A more precise statement of this theorem and its proof can be found in Appendix~\ref{app:convex_hull}. We also note that this theorem has appeared before in the mathematical literature \cite{fenchel1951differential}. 

We now use this general method to find a family of pulses yielding $x$-rotations $X_\theta$, which when combined with virtual $z$-rotations and/or phase ramping can give any single-qubit gate.  The pulses in Fig. \ref{tilted curve}c are derived from tangent curves comprised of "tilted circles", as shown in Fig. \ref{tilted curve}(a) and \ref{tilted curve}(b). The curve goes from $\vec{T}_0 = \qty\big(\cos(\theta/2),0,\sin(\theta/2))^T$ to $\vec{T}_f = \qty\big(\cos(-\theta/2),0,\sin(-\theta/2))^T$ along a great circle arc. However it goes around two smaller circles before and after to cancel the area swept out by the great circle arc. The upper and lower circles are normal to the vectors $\hat{n}_0 = \qty(0, -\sin\alpha, \cos\alpha)^T$ and $\hat{n}_f = \qty(0, -\sin\alpha, -\cos\alpha)^T$, respectively. The sign of the normal vectors are chosen so that the area contribution from the circle points along $\hat{n}$:
\begin{equation}
  \vec{A}_{\text{circle}} = \pi r_{\text{circle}}^2 \hat{n} = \pi \sin^2 \gamma~\hat{n},
\end{equation}
where $\gamma$ is defined as the angle between $\vec{T}_0$ and $\hat{n}_0$ (and the angle between $\vec{T}_f$ and $\hat{n}_f$):
\begin{equation}
  \cos\gamma = \vec{T}_0 \cdot \hat{n}_0 = \vec{T}_f \cdot \hat{n}_f = \sin\frac{\theta}{2} \cos\alpha.
\end{equation}
The area contribution of the arc from $\vec{T}_0$ to $\vec{T}_f$ is $\vec{A}_{\text{arc}} = \theta/2 \hat{y}$, and the total area swept out is
\begin{align}
  \vec{A} &= \frac{\theta}{2}\hat{y} + \pi \sin^2 \gamma~\qty(\hat{n}_0 + \hat{n}_f) \nonumber\\
  &= \qty(\frac{\theta}{2} - 2\pi\sin\alpha\qty(1-\sin[2](\frac{\theta}{2})\cos^2 \alpha))\hat{y}.
\end{align}
In order to cancel driving error, we require $\vec{A} = 0$, which gives an implicit equation defining $\alpha(\theta)$: $2\pi\sin\alpha\qty(1-\sin[2](\theta/2)\cos^2 \alpha) = \theta/2$. We also see that the origin is contained in the convex hull of this curve, and so it can be reparameterized to give a closed space curve. In Fig.~\ref{tilted curve}, we validate the inclusion of the origin in the convex hull of the tangent curve and present the infidelity of the resulting evolution. The area of low infidelity is the largest among all our results, which emphasizes the fact that the degree of error cancellation affects the rate at which the quality of the gate degrades with increasing noise strength and not its absolute fidelity.

\section{Conclusion}\label{sec:conclusion}
In conclusion, we presented the description of the geometric control in the language of the (SCQC) formalism defined in the error interaction picture. Utilizing the Magnus expansion, we can transfer the noise suppression problem entirely to a geometric space curve where each order of error is associated with a respective geometric property. Specifically, a zero total area of the tangent curve leads to cancellation of first-order multiplicative error and a closed curve reduces the contribution of a static error. The relationship to the dynamical phase is subsequently revealed and manifests the fact that its cancellation does not necessarily imply total tolerance to (first-order) errors in the driving field. The validation of our theoretical results are supported by simulation of three types of curves: trigonometric "parity curves", "Bessel curves" and the "tilted circles" approach, which are curves satisfying both geometric conditions. Analysis of the infidelity plots reveals the significantly greater performance of such gates compared to gates that don’t satisfy either robustness conditions. Mediated by the gauge freedom provided by the torsion, any desired phase accumulation does not degrade the level of robustness of this geometric scheme. 

\section*{Acknowledgments}

This work is supported by the Office of Naval Research (grant no. N00014-21-1-2629), the Army Research Office (grant no. W911NF-17-0287), and the Department of Energy (grant no. DE-SC0022389).

\appendix

\section{Signed curvature and inflection points}
\label{Curve to fields mapping}
The introduction of the moving frame allows us to uniquely map the control fields to the Frenet-Serret vectors, in an interval free of inflection points, essentially $\kappa(t) \neq 0$. 
In general, Eq. \eqref{torsion-field-eq} implicitly assumes the differentiability of $\Omega(t)$, despite the fact that $\dot\Omega(t)$ does not appear explicitly. When inflection points (points where the curvature vanishes) arise, the curvature function contains cusps at these points; therefore a gauge transformation allows us to use a generalized version of the curvature which changes sign at every inflection point in the interval of evolution, ensuring the continuity of the normal vector \cite{Hu2011}. If $t_i^*$ are points of zero curvature, then, the driving field is given by:
\begin{align}
    \Omega(t) = \kappa(t)\sum_{i=1}^N (-1)^i\theta(t-t_i)\theta(t_{i+1}-t).
\end{align}
Essentially, this is a form of signed curvature. Similarly, the phase field exhibits a $\pi$ discontinuity at such points, yielding the same behavior for the fields along $\sigma_x$ and $\sigma_y$.
The same technique was introduced in the context of the SCQC formalism in Ref.~\cite{Zhuang_2022}.

\section{Convex Hull Theorem}
\label{app:convex_hull}
Here we give a more precise statement and proof of Theorem \ref{thm:conv_hull} from the main text:

\begin{theorem}
    A tangent curve $\vec{T}(s)$ can generate a closed space curve if and only if the convex hull of $\vec{T}(s)$ contains the origin. Additionally in order for $\kappa(t)$ to be finite at all times (no delta function pulses), the origin must be in the interior of the convex hull of $\vec{T}(s)$.
    
\end{theorem}

The convex hull of $\vec{T}(s)$ is the set of vectors

\begin{equation}
  \mathcal{C} = \qty{\int_0^{s_f}\dd{s} \lambda(s) \vec{T}(s): \lambda(s) \geq 0, \int_0^{s_f}\dd{s} \lambda(s) = 1}.
\end{equation}

Given some particular parameterization $s(t)$, we can write $\vec{R}(T)$ as

\begin{align}
  \vec{R}(T)& = \int_0^T \dd{t} \vec{T}(t) = \int_0^S \dd{s} \dv{t}{s} \vec{T}(s) \nonumber\\
  &= T \qty(\int_0^S \dd{s} \frac{1}{T \kappa(s)} \vec{T}(s)),
\end{align}

where $S \equiv s(T)$. We can therefore identify $\vec{R}(T)/T$ as a point in $\mathcal{C}$, with $\lambda(s) = 1/T \kappa(s)$; clearly $1/T \kappa \geq 0$ since $\kappa \geq 0$, and
\begin{align}
  \int_0^S \dd{s} \frac{1}{T \kappa(s)} = \frac{1}{T} \int_0^S \dd{s} \dv{t}{s} = \frac{1}{T} \int_0^T \dd{t} = 1.
\end{align}
Thus in order for the space curve to be closed, i.e. $\vec{R}(T) = 0$, there must be some choice of $\lambda(s) = 1/T \kappa(s)$ such that $\int_0^{s_f}\dd{s} \lambda(s) \vec{T}(s) = 0$, i.e. the origin must be in the convex hull of $\vec{T}$.

Additionally, if $\kappa$ is finite at all times, then $\lambda(s) = 1/T \kappa(s) > 0$. The points in $\mathcal{C}$ with $\lambda(s) > 0$ for all $s$ are points in the interior of $\mathcal{C}$.

%


\end{document}